\documentclass[journal, a4paper]{IEEEtran}
\usepackage[left=0.70in,right=0.70in,top=0.85in,bottom=1.5in]{geometry}

\usepackage{amssymb}
\usepackage{amsmath}
\usepackage{textcomp}
\usepackage{amsfonts}
\usepackage{amsthm}
\usepackage{cite}
\usepackage{pdflscape}
\usepackage{tabularx}
\usepackage{paralist}
\usepackage{algorithm}
\usepackage[noend]{algpseudocode}

\usepackage{enumitem}
\usepackage{pdflscape}
\usepackage{multicol}
\usepackage{multirow}
\usepackage{url}
\usepackage{graphicx}
\graphicspath{ {Images/} }
\ifCLASSINFOpdf
\else
 \fi

\begin{document}

\pagenumbering{gobble}

\title{A Review of Routing Protocol Selection for Wireless Sensor Networks in Smart Cities}

\author{\IEEEauthorblockN{Mohsin Khalil$^*$, Ammar Khalid$^*$, Farid Ullah Khan$^\#$, Akmal Shabbir$^*$}\\
\IEEEauthorblockA{$^*$National University of Sciences and Technology (NUST), Islamabad, Pakistan \\
$^\#$Air University, Islamabad, Pakistan \\
}
}

\maketitle

\begin{abstract}

Today, the advancements in urban technology have transformed into the concept of smart cities. These smart cities are envisioned to be heavily dependent on wireless sensor networks and internet of things. In this context, a number of routing protocols have been proposed in literature for use in sensor networks. We articulate on why these routing protocols need to be segregated on the basis of their operational mechanism and utility, so that selection of these protocols results in network longevity and improved performance. We classify these protocols in four categories in terms of topology incognizant, data centric, location assisted and mobility based protocols. We identify the prevailing open issues to make space for more productive research and propose how these categories may be useful in terms of their operational utility.

\end{abstract}


\section{Introduction}

The concept of a smart city is used globally with various definitions and nomenclatures. In most of the cases, the terminology is generally coined to represent the advancements in urban technology through systems integration \cite{albino2015smart}. As per the United Nations statistics, it is estimated that by year 2050, around 68\% of the world population would be living in urban areas \cite{wup18}. This situation would invariably result in the challenges of urbanization, which include (and are not limited to) waste management, air pollution, traffic management, human health and disaster management --- an area under active research of late \cite{pervez2018wireless}. Although there is an increased concern worldwide regarding the performance enhancement of networks \cite{khalil2018exploration} while also keeping an eye on the rural-urban digital divide due to technological advancements in urban areas \cite{khalil2017feasibility}, however, the search for innovative solutions to ensure more efficient urban dynamics for conservation of resources and optimal use of technology is still a work in progress.

\vspace{0.1cm}

The core of infrastructure in smart cities is sensing \cite{ahvenniemi2017differences}. The appropriate use of effective sensors with ubiquitous range and efficient communication is the need of the hour to ensure optimal resource utilization. Such an approach can only yield fruitful results when the number of sensors is significantly high and are mutually connected to each other for exchange of information. This calls for the formulation of a communication infrastructure and a mechanism for data aggregation and processing. Therefore, various approaches which may be considered in case of such an infrastructure are Internet of Things, Cloud of Things and Wireless Sensor Networks (WSNs).

\vspace{0.1cm}

Wireless sensor networks comprise of battery powered sensor nodes, which essentially means that they are energy constrained. In smart cities, number of these nodes would be invariably high, therefore, the possibility of replacement or recharging of these nodes becomes scarce. Therefore, it becomes imperative to identify the suitability of WSN protocols vis-a-vis given scenario and use-case so that it may serve as a guideline in terms of operational utility.

\subsection{Contributions of this Paper}

In this work, we first motivate the need of categorizing WSN protocols based on their operational mechanism and utility. Furthermore, we classify the protocols in four categories in lieu of their use cases in smart cities. Table I represents the comparison of our survey in terms of various aspects with existing surveys on WSNs. Moreover, we also highlight the smart city scenarios where this classification would help in appropriate selection of routing protocols based on the features of the proposed categories.

\subsection{Organization of the Paper}

This paper is divided into five sections. Section II enunciates the factors which are responsible for loss of data in WSNs in practical situations. In Section III, we classify the recent WSN protocols in four categories to enable the reader in terms of apt selection based on varying use cases in smart cities. Section IV highlights the open issues and active research areas, followed by conclusion of the paper in Section V.

\renewcommand{\arraystretch}{1.50}
\begin{table*}
\caption{Comparison with Existing Surveys on Wireless Sensor Networks} 
\centering
\begin{tabular}{|c|c|c|c|}
  \hline

  {\begin{tabular}{cc} \textbf{Classification Criteria} \\ \textbf{(Orientation)}\end{tabular}} & {\begin{tabular}{c} \textbf{Contributed by} \end{tabular}} & {\begin{tabular}{c} \textbf{Year} \end{tabular}} & {\begin{tabular}{c} \textbf{Categories} \end{tabular}}\\
  \hline

  Energy saving & {\begin{tabular}{cc} Rault et al. \\ \cite{rault2014energy} \end{tabular}}
  & 2014 &
  {\begin{tabular}{cc} Radio Optimization, Data Reduction, Sleep / Wakeup Schemes, \\ Energy-efficient Routing, Battery Depletion \end{tabular}} \\
  \hline

  {\begin{tabular}{cc} Protocol \\ Composition / Techniques \end{tabular}} & {\begin{tabular}{cc} Deif et al. \\ \cite{deif2014classification} \end{tabular}}
  & 2014 &
  {\begin{tabular}{cc} Genetic Algorithms, Computational Geometry, Artificial Potential Fields, \\ Particle Swarm optimization \end{tabular}} \\
  \hline

  Environment & {\begin{tabular}{cc} Fadel et al. \\ \cite{fadel2015survey} \end{tabular}}
  & 2015 &
  Terrestrial, Underground, Underwater, Multimedia, Mobile\\
  \hline

  Energy Harvesting & {\begin{tabular}{cc} Shaikh et al. \\ \cite{shaikh2016energy} \end{tabular}}
  & 2016 &
  RF based, Solar based, Thermal based, Flow based, Mechanical based, Human based\\
  \hline

  Metric Based & {\begin{tabular}{cc} Yuan et al. \\ \cite{yuan2017instrumenting} \end{tabular}}
  & 2017 &
  Node centric, Hop-centric, Path centric, End-to-end, Network centric\\
  \hline

  Hierarchy & {\begin{tabular}{cc} Sabor et al. \\ \cite{sabor2017comprehensive} \end{tabular}}
  & 2017 &
  Classical based routing protocols, Optimization based routing protocols \\
  \hline


  {\begin{tabular}{cc} \textbf{Operational Mechanism} \\ \textbf{in Smart Cities}\end{tabular}} & {\begin{tabular}{c} \textbf{Our Survey} \end{tabular}} & {\begin{tabular}{c} \textbf{2018} \end{tabular}} & {\begin{tabular}{c} \textbf{Topology Incognizant, Data Centric, Location Assisted, Mobility Based 
  } \end{tabular}}\\
  \hline

  \end{tabular}
\label{Comparison with Existing Surveys on Wireless Sensor Networks}
\end{table*}

\section{Motivation for Operation-based Classification of Routing Protocols}
\vspace{0.1cm}
Although the sensor nodes are resource constrained, however the aggregated energy in the entire network is sufficient for the desired goal. The sensor nodes are powered, generally have limited battery and are deployed in diverse environments, so it is important to optimize the performance of the network since it is not possible to replace or charge the batteries. The sensor nodes are deployed randomly and have this characteristic of autonomous configuration for transforming into a network. The deployment scenario of wireless sensor networks is bound to change the network topology due to battery drainage of sensor nodes, mobility or channel fading. Hence, their classification based on operational mechanism would assist in appropriate protocol selection based on the use case, which might be helpful in minimizing energy wastage as a result of data communication within smart cities. This will be particularly useful in following aspects:

\subsection{Node Deployment}

Sensor nodes have to be densely placed in the area of interest in various cases. Selection of various protocols serves the purpose but the threat of battery drainage of sensor nodes is always imminent due to frequent data transfers. So, the node deployment is generally compromized to take care of this aspect, which could have been avoided by the selection of a more suited protocol.

\subsection{Data Aggregation}

Instead of transmitting similar packets from various nodes, data aggregation aims at combination of data from multiple sources to reduce transmission. This would help in preserving the residual energy of the sensor nodes, thus increasing network longevity.

\subsection{Node Capability}

Various nodes in sensor networks may be assigned roles of sensing, data aggregation or relaying based on the requirement. This causes the node to drain the residual energy more rapidly. Since various nodes have different capabilities, therefore the quick battery depletion makes routing an arduous task.

\subsection{Scalability}

Routing algorithms differ in terms of their support for scalability. Protocols that support scalable architecture are generally computationally complex. Therefore, it is imperative to keep this factor in mind during the selection of routing protocols, as it would be an overkill to use a scalable routing protocol for a small scale application at the cost of significant battery drainage.

\section{Classification of Protocols}
\vspace{0.1cm}
The networking protocols of wireless sensor networks may be classified in four categories based on their operational mechanism and utility in smart cities, details of which are presented in ensuing paragraphs.

\subsection{Topology Incognizant Protocols}

This category comprises of those protocols which do not need comprehensive use of the network topology. These are generally used for data relaying purposes and in various cases, these protocols might not need to update their routing tables. Yao et al. \cite{yao2015edal} have presented a data collection protocol EDAL for heterogeneous wireless sensor networks. The concept is based on mapping the results from Open Vehicle Routing (OVR) problems onto wireless sensor networks, which renders the formulation of the original algorithm as an NP-hard problem. The issue was solved with the help of a centralized heuristic for reducing the overhead, so that the computational complexities may be reduced to a certain extent. However, limitation in this case is of scalability, as it can only
be addressed by employing a distributed heuristic algorithm in order to support large scale network operations. Therefore, it may only be used for relay purposes in smart city deployments of WSNs.

Some works have gauged the performance of routing protocol using optimization techniques, however they all may not be classified in same category as operational mechanism would vary with the solution proposed for the optimization problem. In NSGA-II protocol \cite{bara2015multi}, an evolutionary algorithm has been developed to solve the problem. With each iteration, a set of solutions is obtained which allows the sink freedom to choose for making clustered routes within the network. Although detection accuracy could also be considered as a performance metric instead of coverage percentage, however this aspect has not been addressed in this approach, which renders this protocol suitable to be categorized as topology incognizant. Moreover, the idea of multihop relay on the principle of opportunistic routing has also been presented in literature, which is useful in selection of appropriate relay node in the network \cite{luo2015opportunistic}. This approach makes use of the distance between sensor node and sink and determines the optimal transmission distance. Subsequently, an energy optimal strategy is adopted based on the optimal distance calculated. Then the selection of relay nodes takes place on the basis of residual node energy and their distances to the sink. The execution of the algorithm results in optimal selection of relay nodes, which results in network longevity and energy efficiency.

Gupta et al. in \cite{gupta2016genetic} have considered the problem of determining the minimum number of relay nodes and their placement for k-connectivity in sensor networks. Since their formulated problem was NP-hard, therefore genetic based algorithm and greedy approach were considered for this purpose. The simulation results reveal that genetic based approach resulted in earmarking lesser number of relay nodes as compared to greedy approach.

\subsection{Data Centric Protocols}

Due to a large number of nodes in Wireless Sensor networks, addressing might not be a preferred option in certain situations. Therefore, every node tries to transmit data towards the sink, which may cause data redundancies. In this context, various routing mechanisms have been proposed in literature in which a certain number of nodes are selected based on a defined criteria and thus data is node-wise aggregated.

For data aggregation, some protocols utilize amalgamated routing metrics. These metrics are based on transmission count, and problems may arise in case of bursty traffic and data impulses. Hence, these protocols have to be essentially multipath for the purpose of congestion avoidance, and CA-RPL is one such example \cite{tang2016toward}. Owing to data aggregation at specified nodes, the average delay by the use of proposed CA-RPL protocol has been observed to decrease by 30\% as compared to conventional approaches. Moreover, wireless multi-sensor networks have also been tested which integrate the data from various sensors and statistically analyze the data \cite{csaji2017wireless}. The metrics such as air quality and traffic are communicated to a central database for subsequent processing.

Various studies indicate that the communication process consumes much more power than the processing and sensing functionalities. Same logic has also been utilized by Jose et al. in \cite{jose2015mobile}, where they proposed a Mobile Sink Assisted algorithm (MSA) for data collection after culmination of three different phases instead of gathering it on a continuous basis. The simulation results showed a considerable improvement in terms of average packet delay and average energy consumption was optimized over longer time samples.

The holistic aggregation of data in case of wireless sensor networks is not a feasible option. In this regard, approximate holistic aggregation has been coined for the purpose of energy savings, and the corresponding algorithms make use of uniform sampling \cite{li2017approximate}. The work presents four mathematical estimators for aggregation process and mathematical methods were utilized for determining the sample size of these estimators. The algorithms are vindicated by the results which indicate accuracy and energy efficiency by selection of appropriate nodes for data aggregation.

A few routing protocols support phase-wise data aggregation in WSNs \cite{harb2017distance}. The phases are aimed to minimize the data transmission from sensors as well as cluster heads. The protocol identifies the similarities among the measurements collected by the sensors in the initial phase. In the second phase, correlation techniques based on distance function are applied among the datasets obtained in consecutive intervals. The experimental results indicate a decrease in data redundancy while ensuring data preservation and energy savings. An alternative to phase-wise aggregation has been presented in \cite{abbasi2016toward}. The algorithm formulates a collection tree by uniform selection of collector nodes. The collector nodes aggregate compressive sensing measurements from client nodes, which are then communicated to the sink. Furthermore, this approach is dovetailed with the clustering methods to obtain energy efficiency, and the results indicate that up to 53\% energy savings could be achieved this way.

\subsection{Location Assisted Protocols}

Location assisted protocols facilitate routing on the basis of information regarding the geographical location of the static sensor nodes and then determine which path is the most energy efficient in order to ensure the longevity of the network. This may be accomplished by use of  GPS or localization algorithms. The location table of each node is updated periodically with respect to its active / passive state. It is pertinent to mention that the protocols for location-assisted mobile nodes have been included in mobility based protocols, which would be discussed later.

Location assisted protocols employ variety of techniques such as distance estimation mechanisms, position estimation of nodes and range estimation methods. In case of distance estimation mechanisms, the use of weighted distance-vector hop algorithm has been proposed based on received signal strength, which calculates the average hop-distance among the nodes via path loss models and corrections are applied through the use of GPS \cite{xiao2017rssi}. The results indicate that the use of location assisted approach outperforms conventional algorithms in use. Moreover, instead of counting the number of hops that are encountered between the nodes, Sanchez et al. have proposed weighted distance vector algorithm fused with weighted hyperbolic positioning for determining location of nodes \cite{mass2017weighted}. Although this combination causes the computation complexity to increase somewhat, however, it results in greater accuracy and precision.

Various information fusion mechanisms are utilized in location discovery algorithms. In a recent study \cite{abu2018localised}, various fusion techniques have been summarized including Bayesian inference, maximum likelihood, Kalman filtering and moving average filters. Information fusion can assist the localization algorithms in leading and support roles. In the former, the information fusion techniques concurrently guide the location discovery and fusion processes, whereas the later focuses on assisting the location discovery by acting as second fiddle to the localization algorithms.

Location assisted protocols can also make use of varying the transmission power based on the proximity of communicating nodes. Sammut et al. \cite{sammut2015location} have proposed a similar algorithm which involves the use of a greedy forwarding mechanism. It works on adjusting the transmission power in accordance with the distance between the nodes. As a consequence, energy savings of 33\% while maintaining 99\% packet delivery was observed.

Another approach utilized in \cite{singh2016modified} classifies the network area in four regions. The gateway node is located at the center of sensing region. By the use of GPS, the distance between the nodes is determined prior to data transmission. In the first round, cluster heads are earmarked in each region. However, in the subsequent rounds, the cluster heads are selected on the basis of distance between the nodes and probability of residual energy within the nodes. This technique curtails the energy consumption of the nodes which is manifested in terms of their lifetime. Another study indicates that since the range-based localization is not practically feasible due to hardware constraints, therefore range-free localization techniques may be adopted and position error as a result of hop-distance estimation may be corrected by the use of genetic algorithm \cite{peng2015improved}.

\subsection{Mobility Based Protocols}

Wireless sensor networks are deployed in diverse environments, so it is necessary to maximize their performance whatever environment and conditions they are exposed. Some of the scenarios may require the nodes to continuously remain moving, thus generic protocols might not be efficient enough to meet the goal. This would essentially require to calculate the optimal route for data transportation from one node to the other. The route calculation has to be dynamic to cater for the mobile nodes that might be in range one moment and out of range at other instants of time.

Recently, use of smart dumpsters for effective waste management in smart cities has been proposed \cite{idwan2016smart}. The dumpsters consist of waste sensors which determine the garbage level and communicate with a control unit. The control unit then communicates with trucks for waste collection which dynamically optimize their paths based on the indication of the sensor nodes reflecting high garbage levels. Additionally, the concept of a smart parking system has also been proposed, whose operational mechanism renders it to be categorized under the umbrella of mobility based protocols. It makes use of a self-organizing algorithm that focuses on energy saving of the sensor nodes during the data communication phase \cite{hilmani2018designing}. Moreover, it may also be utilized to assist the drivers in the city towards the closest vacant parking based on optimal path calculation on the fly.

Jamil et al. \cite{jamil2015smart} have suggested the use of wireless sensors on automobiles for continuously determining the air pollution metrics. The proposed methodology is based on the direct communication of mobile nodes on the vehicles with the static nodes in the smart city for transporting the air pollution data. This effectively results in time management and energy savings are obtained due to the involvement of only the relevant nodes, which results in the formulation of best route towards the data hub. Thus, the energy saving ensures the long battery life of sensor nodes, which translates into the network longevity as well. Moreover, Hammoudeh et al. have presented a Nutrient-flow based protocol NDC for adaptive routing in wireless sensor networks, which is based on various QoS guarantees to improve the overall network performance \cite{hammoudeh2015adaptive}. The optimization tool was developed to balance the available network resources with the application related user priorities serving as the constraints of the optimization problem. Subsequently, a new algorithm was dovetailed along with the optimization tool for load balancing purposes resulting due to mobile nature of the nodes. Simulation results revealed that the setup messages were reduced by 60\% and data delivery ratio was improved by 0.98\% with the use of NDC against other similar schemes. Moreover, a different approach hints at consideration of mobile sinks because the nodes near the sink are likely to drain quickly as compared to other nodes owing to traffic concentration \cite{tunca2015ring}.

\section{Open Issues and Active Research Areas}
\vspace{0.1cm}
In spite of vast contributions to this field, there are still some open issues that require particular attention. Some of the dominant areas in the regard are listed below.

\renewcommand{\arraystretch}{1.50}
\begin{table*}
\caption{Protocol Classification in Wireless Sensor Networks from an Operation oriented viewpoint} 
\centering
\begin{tabular}{|c|c|c|}
  \hline

  {\begin{tabular}{c} \textbf{Categories} \end{tabular}} & {\begin{tabular}{r} \textbf{Features} \end{tabular}} & {\begin{tabular}{c} \textbf{Operational Utility} \end{tabular}}\\
  \hline

  Topology Incognizant & {\begin{tabular}{ll} - Topology information is not required \\ - May be used for relaying data \end{tabular}}
  &
  Disaster site recovery, Urban internet \\
  \hline

  {\begin{tabular}{c} Data Centric \end{tabular}} & {\begin{tabular}{ll} - Data aggregation on selective nodes \\ - Avoidance in data redundancy \end{tabular}}
  &
  Healthcare, Emergency Response \\
  \hline

  Location Assisted & {\begin{tabular}{ll} - Location estimation of the static nodes \\ - Node status indication in location table \end{tabular}}
  &
  {\begin{tabular}{cc} Temperature Monitoring, Waste Management, Smart Grid, \\ Gas monitoring, Water distribution monitoring \end{tabular}} \\
  \hline

  Mobility Based & {\begin{tabular}{ll} - Specific to mobile nodes \\ - Dynamic calculation of optimal routes \end{tabular}}
  &
  Air Pollution, Transportation Systems\\
  \hline

  \end{tabular}
\label{Protocol Classification in Wireless Sensor Networks from an operation oriented viewpoint}
\end{table*}

\vspace{0.1cm}
\emph{Maximizing Power Efficiency: } One of the main research areas is to maximize the power efficiency of sensor nodes which would yield improved performance. Since the wireless sensor networks are essentially power limited, so achieving efficiency in power utilization is the obvious goal.

\vspace{0.1cm}
\emph{Multipath Routing: } Routing within wireless sensor networks has to be essentially multipath because the circumstances and the conditions of the deployment scenarios may vary with time.

\vspace{0.1cm}
\emph{Characteristics of System Components: } Two key terms, durability and robustness, summarize the problem. It is still a challenge to design such sensors which can withstand the challenging weather situations such as rain as well as high temperatures since the terrain features of a natural calamity or man-made disaster cannot be predicted. Thus the use of wireless sensor networks for emergency response situation calls for improvement in this regard.

\vspace{0.1cm}
\emph{Mobility Support: } Due to the evolution of networks and the ever-increasing requirements, the increase in mobility support of the sensor nodes within wireless sensor networks needs to be taken care of so that it does not behave as a bottleneck. Some of the nodes within a wireless sensor network would be essentially mobile and this factor may hamper the network performance if not addressed properly.

\vspace{0.1cm}
\emph{Security Issues: } Security is a major concern for wireless sensor networks which needs due emphasis and refinement. A hacker can gain access to sensor data and can disturb the functionality of whole network by overwriting wrong information.

\section{Conclusion}

As a successful technology to serve in urban areas, wireless sensor networks still have a lot of ground to cover because energy constrained sensors with compromised processing ability need to effectively operate in diverse environment of multiple applications. To use small, lightweight and portable sensors for multitasking, we need appropriate protocols which enables sensors to work properly by avoiding overkill so that energy levels of sensor nodes are not depleted. Therefore, we opine that efficient operation may be achieved by rendering some of the sensor nodes for data aggregation in one situation, whereas another scenario would call for a few nodes to act as data relays instead of aggregators. Similarly, the selection of routing protocols would vary depending upon whether the intended operation is location assisted or mobility based. We have discussed and categorized routing protocols in this paper which may serve as a selection guideline in terms of their operational mechanism and utility. Table II presents these categories along with their prominent features and operational utility for practical scenarios in smart cities. Furthermore, we have also articulated some open research areas that require further exploration.

\bibliographystyle{IEEEtran}
\bibliography{Survey_Paper}

\end{document}